\documentstyle[12pt,aaspp4]{article}
\begin{document}

\title{A model independent lower limit on the number of Gamma Ray Burst hosts
from repeater statistics}
\author{Anupam Singh\altaffilmark{1} and Mark Srednicki\altaffilmark{2}}
\affil{Department of Physics, University of California, Santa Barbara,
CA 93106}
\altaffiltext{1}{E-mail: singh@solkar.physics.ucsb.edu}
\altaffiltext{2}{E-mail: mark@tpau.physics.ucsb.edu}

\begin{abstract}
We present a general statistical analysis of Gamma Ray Bursts embedded in a 
host population.  If no host generates more than one observed burst, then we 
show that there is a model independent lower bound on the number of hosts, $H$,
of the form $H > c B^2$, where $B$ is the number of observed bursts, and $c$ is
a constant of order one which depends on the confidence level (CL) attached to
the bound. An analysis by Tegmark et al.~(1996) shows that 
the BATSE 3B catalog of 1122 bursts is consistent with no repeaters
being present, and assuming that this is indeed the case, our result
implies a host population with at least $H = 1.2\times 10^6$ members.
Without the explicit assumption of no repeaters,
a Bayesian analysis based on the results of Tegmark et al.~(1996) can be 
performed which gives the weaker bound of $H>1.7\times 10^5$ at the 90\% CL.  
In the light of the non-detection of identifiable hosts in the small 
error-boxes associated with transient counterparts to GRBs,
this result gives a model independent lower bound to the number of
any rare or exotic hosts. If in fact GRBs 
are found to be associated with a particular sub-class of galaxies, 
then an analysis along the lines presented here can be used to place
a lower bound on the fraction of galaxies in this sub-class.  
Another possibility is to treat galaxy clusters (rather than individual
galaxies) as the host population, provided that the angular size
of each cluster considered is less than the resolution of the detector.
Finally, if repeaters are ever detected in a statistically significant manner,
this analysis can be readily adapted to find upper and lower limits on $H$.  
\end{abstract}

\subjectheadings{gamma rays: bursts}

\section{Introduction}

Recently there have been exciting developments in the observational
investigation of the nature of Gamma Ray Bursts (GRBs) and the hosts in which
they may originate.  (Several excellent reviews of GRBs are available;
see e.g.~Blaes [1994]; Piran [1995].) For GRB970228, an afterglow was detected
in the X-Ray and optical wavebands (Van Paradijs et al. 1997),
and Metzger et al. (1997) have identified absorption lines in the spectra 
of the optical transient associated with GRB970508.  Their observations put the
transient beyond a redshift of $z = 0.835$.
If confirmed, this would establish that GRBs are cosmological.
If so, then this opens the door to their use as tools for probing
cosmological issues such as structure formation (Piran \& Singh 1997).

The latest observations herald a new era of more
precise, useful information about GRBs. The counterparts in
the X-ray and optical wavebands allow one to get among other
things a more precise location of the bursting source.
Further, one can perform a detailed and intensive survey of the
neighboring region of space to seek out the host population
in which the GRBs are embedded. 
So far, in spite of exhaustive searches, statistical analysis,
and the accompanying debate 
(Fenimore et al. 1993; Larson, McLean \& Becklin 1996; 
Schaefer et al. 1997; Blaes et al. 1997; Kolatt \& Piran 1996;
Hurley et al. 1997),
GRBs have not clearly and unambiguously been associated with
any other well-known population of astronomical objects.
In particular, no known or identifiable
galaxy has been associated with either of the two GRBs for which
transient counterparts have enabled a small error-box.
Regarding repetitions of GRBs from the same host,
while there was an early claim of evidence for this 
by Quashnock \& Lamb (1993), their analysis was called into question 
by Narayan \& Piran (1993). More recently, Tegmark et al. (1996) 
concluded that the data is consistent with no repetition of GRBs.
Thus all of the recent data has not yet resolved the mystery
of the nature of the host population in which GRB sources
are embedded.

It is therefore important to deduce all possible constraints
which can be placed on the host population; model-independent constraints
are particularly valuable.  Obviously, there must be at least as many
hosts as there have been GRBs, assuming no repeaters.
This is clearly not a significant constraint; the total number of galaxies
is about $10^9$, corresponding
to about $10^{-6}$ bursts per galaxy per year.  It is possible, however,
that the host population consists of exotic or otherwise rare cosmological
objects.  In this Letter, we perform a statistical analysis
which demonstrates that, if a total of $B$ bursts have been
seen, and we assume that there are no repeaters, then
the number of hosts $H$ is likely to exceed $cB^2$, where
$c$ is a numerical constant which depends on the desired confidence level
and the precise form of the statistic (e.g., choice of the prior distribution
in a Bayesian analysis), but not on any details of the GRB model.
If we do not make the assumption of no repeaters, but instead 
appropriately weight the possibilities based on observational data,
then the lower bound is reduced to $H_{\rm min} \sim B^2/\langle N_2\rangle$,
where $\langle N_2\rangle$ is the expected number of repeaters in the data.
In this analysis, a ``host'' is any object whose angular size
is much less than the resolution of the detector.

The BATSE data is consistent with no repeaters in the analyzed catalog
of 1122 bursts.  An analysis by Tegmark et al. (1996) shows that
the number of repeaters $N_2$ is less than 10 at the 95\% CL.
If we assume that there are in fact no repeaters in this catalog, then
we can conclude that the the population of hosts from which the GRBs 
may be originating should contain at least
$H = 1.2 \times 10^6$ members.  This implies that about 0.1\% of the 
total number of galaxies could be hosts, if GRBs are embedded in galaxies. 
This could be a 
significant constraint if GRBs are found to be associated with a specific
sub-class of galaxies. This constraint would also be significant
if ongoing searches for potential objects within the error boxes of the
transients fails to turn up identifiable normal galaxies or other well-known
and abundant astronomical objects.  
Furthermore, if repeaters are ever detected in a
statistically significant manner, our analysis can be readily adapted to
find upper and lower limits on the number of hosts $H$.  

We now turn to the statistical
analysis of this problem which allows us to draw the advertised conclusions.

\section{The General Statistical Analysis}

We suppose that there is a population of $H$ hosts, each of which is
producing burst bursts at an individual rate of $\gamma_i$, $i=1,\ldots,H$,
and that each of these hosts has been observed for an amount of time
$t_i$.  Then the number of bursts which have been observed from each
host is expected to be $A_i = \gamma_i t_i$.  Call the actual number of
observed bursts from each host $B_i$, and the total number of observed
bursts $B\equiv\sum_{i=1}^H B_i$. 

We assume that the bursts from each host are uncorrelated, so that the 
probability of observing $B_i$ bursts from host $i$ is given by a 
Poisson distribution
\begin{equation}
P(B_i|A_i) = {A_i^{B_i} \exp(-A_i) \over B_i !} \, .
\label{poisi}
\end{equation}
Then the probability that a total of $B$ bursts are observed is
\begin{eqnarray}
P(B|A,H) &=& \sum_{B_1=0}^B \ldots \sum_{B_H=0}^B 
             \delta_{B_1+\ldots+B_H,B} \; P(B_1|A_1)\ldots P(B_H|A_H) 
\nonumber \\
         &=& {(AH)^B \exp(-AH) \over B!} \, , 
\label{probe}
\end{eqnarray}
where $A$ is the average value $A \equiv (1/H) \sum_{i=1}^H A_i$.

We wish to calculate that probability that $B$ bursts are seen, but that 
there are no repetitions; this is given by
\begin{equation}
P(B,\hbox{no rep}|\{A_i\},H) 
= \sum_{B_1=0}^1 \ldots \sum_{B_H=0}^1
  \delta_{B_1+\ldots+B_H,B} \; P(B_1|A_1)\ldots P(B_H|A_H) 
\label{probno}
\end{equation}
To simplify the calculation, we first treat the idealized situation in which 
$A_i=A$ for each host.  In this case, eq.~(\ref{probno}) becomes
\begin{equation}
P(B,\hbox{no rep}|A,H) = {H!\over B!(H-B)!}\,A^B \exp(-AH) \, ,
\label{probno2}
\end{equation}
The probability that no repetitions are seen, given that $B$ bursts are
observed, is given by the ratio
\begin{eqnarray}
P(\hbox{no rep}|B,A,H) &=& {P(B,\hbox{no rep}|A,H) \over P(B|A,H)} 
\nonumber \\
                       &=& {H! \over H^B (H-B)!} \, ,
\label{ratio}
\end{eqnarray}
which we see is independent of $A$.  Assuming $H \gg B \gg 1$,
and using Stirling's formula for the factorials, we get 
\begin{equation}
P(\hbox{no rep}|B,A,H) = \exp(-B^2/2H) \, .
\label{ratio2}
\end{equation}
This is our key result.  It tells us that we are unlikely to have seen
no repetitions among $B$ bursts if $H$ is significantly smaller that $B^2$.
This can be made more precise by using a Bayesian analysis; see,
e.g., Loredo and Wasserman (1995).  
Assuming that no repetitions are seen, the probability 
that the number of hosts is between $H$ and $H+dH$
is proportional to the likelihood $P(\hbox{no rep}|B,A,H)$ 
times a prior distribution;
if we make the standard scale-invariant choice $dH/H$ with an upper cutoff
at $H_{\rm max}=10^9$, we find that $H>B^2$ with a confidence level of 
approximately 90\% (the exact value depends weakly on $B$ and $H_{\rm max}$).

If some hosts generate repeat bursts, the analysis above is changed somewhat.
Let $N_\nu$ be the number of hosts which generated $\nu$ bursts each;
then $H=N_0+N_1+N_2+\ldots$ and $B=N_1+2N_2+3N_3+\ldots$.  We assume
$N_0\gg N_1 \gg N_{2,3,\ldots}$.  Then eq.~(\ref{ratio}) is replaced with
\begin{eqnarray}
P(N_2,N_3,\ldots|B,A,H) &=&
{H! \, B! \, H^{-B} \over (0!)^{N_0} N_0! \ldots (\nu!)^{N_\nu} N_\nu ! \ldots} 
\nonumber \\
&=& {(B^2/2H)^{N_2} \over N_2 !} \exp(-B^2/2H) \exp(-2 N_2^2/B)  
    \delta_{N_3 0} \delta_{N_4 0} \ldots \, .
\label{newratio}
\end{eqnarray}
Terms with $N_\nu\ne 0$ for $\nu\ge 3$ are suppressed by 
at least one factor of $B/H$.  Henceforth, we set $N_3=N_4=\ldots=0$.

Let us now consider what can be deduced from our formulas when they
are applied to the BATSE 3B catalog of $B=1122$ bursts.
First we will examine the results of Tegmark et al (1996) concerning
repeaters.  To gain an analytic understanding of these results,
we will treat the simplified case in which the beam-width $\sigma$ 
is the same for every burst; then their statistic $R$ becomes
\begin{equation}
R = {1\over 2B}\sum_{i=1}^B\sum_{i=1}^{j-1}
               \exp(-\theta_{ij}^2/4\sigma^2) \,,
\label{r}
\end{equation}
where $\theta_{ij}$ is the angle on the sky between bursts $i$ and $j$.

We now compute the expected values of $R$ and $R^2$, assuming that 
the hosts are uniformly distributed on the sky, 
that the angular size of each host is much less than $\sigma$,
and that each the observed location of each burst is displaced 
from its true value by a random
angle selected from a gaussian distribution of width $\sigma$.  
(The latter two points are relevant only for bursts from repeaters.)
In the flat-sky approximation ($\sigma \ll 1$), we find
\begin{eqnarray}
\langle R\rangle &=& {\textstyle{\frac14}}B\sigma^2 + 
                     {\textstyle{\frac18}}f
\label{rav} \\
(\Delta R)^2 &\equiv& \langle R^2\rangle - \langle R\rangle^2 
\nonumber \\
  &=& {\textstyle{\frac1{16}}}\sigma^2 + {\textstyle{\frac1{96}}}B^{-1}f \, ,
\label{dr}
\end{eqnarray}
where $f \equiv 2N_2/B$ is the fraction of bursts which came from repeaters.
If each host has an angular size $\phi$, then the coefficient of $f$ in
eq.~(\ref{rav}) is reduced; if the burst probability across a host
has a gaussian profile, then the reduction factor is 
$\sigma^2/(\sigma^2 + \phi^2)$.  Not including such a factor in our
subsequent analysis implies that each of our hypothetical hosts has an
angular size $\phi$ which is much less than $\sigma$.

Eqs.~(\ref{rav},\ref{dr}) can be understood heuristically.
Roughly speaking, $\langle R\rangle$ is the expected number of bursts 
which lie within an angular distance $\sigma$ of any one particular burst.
There is a contribution to $\langle R\rangle$ due to bursts from 
different hosts; this contribution is of order $B\sigma^2$.
There is a second contribution if the particular burst in question
is a repeater; the probability that this is the case is of order $f$.
This explains the structure of eq.~(\ref{rav}).  To understand the
variance $(\Delta R)^2$, we note that the $B\sigma^2$ term in 
$\langle R\rangle$ comes from a Poisson process, so its variance
is also $B\sigma^2$.  However, averaging over the particular burst
in question reduces the variance by a factor of $B$.  The term of
order $f$ in $(\Delta R)^2$, which in practice is not important,
can be similarly understood.

The probability distribution for $R$ should be approximately gaussian,
\begin{equation}
P(R|N_2,B) = {1\over\sqrt{2\pi}\Delta R}
             \exp[-(R-\langle R\rangle)^2/2(\Delta R)^2] \;,
\label{probr}
\end{equation}
where $R_{\rm av}$ and $\Delta R$ are given in terms of $N_2 = fB/2$ and $B$
by eqs.~(\ref{rav},\ref{dr}).  For BATSE, in which
each burst has its own value of $\sigma$, and in which sky coverage is
not uniform, eqs.~(\ref{rav},\ref{dr}) should be approximately correct, 
but possibly with different numerical coefficients.  
For this case, we turn to the Monte Carlo results of Tegmark et al.~(1996).
Their Figure~1 shows $\int_0^R P(r)dr$
vs.~$R$ for several different values of $f$.  From this, we can read off
$\langle R\rangle = 0.316 + 0.15\,f$ and $\Delta R = 0.0068$ for $f=0$.  
This is consistent with our eqs.~(\ref{rav},\ref{dr})
with a value of $\sigma\sim 2^\circ$.
(We neglect the $f$ dependence of $\Delta R$,
which is difficult to extract accurately, since it will not be important
to our conclusions.)
The experimental value of $R$ for BATSE is $R_{\rm exp}=0.308$.
Assuming that there are in fact no repeaters, then the
probability that the BATSE data would yield a value of $R$ which
is less than $R_{\rm exp}$ is 0.11.  Eq.~(\ref{probr}) is also consistent
with the results of Tegmark et al. (1996) that
$f<0.018$ at 95\% CL and $f < 0.049$ at the 99\% CL.

We now combine eq.~(\ref{newratio}) and eq.~(\ref{probr}) to get
\begin{equation}
P(R_{\rm exp}|B,A,H) =
\sum_{N_2=0}^{B/2} P(R_{\rm exp}|N_2,B)  P(N_2|B,A,H) \;.
\label{data}
\end{equation}
Again performing a Bayesian analysis for $H$ with the scale-invariant
prior distribution $dH/H$ and an upper cutoff of $H_{\rm max}=10^9$,
we find $H>1.7 \times 10^5$ with a CL of 91\%; this is smaller than $B^2$
by a factor of $7.4$.  This factor is roughly the expected number
of repeaters in a Bayesian analysis, 
\begin{equation}
\langle N_2\rangle = {\sum_{N_2=0}^\infty P(R_{\rm exp}|N_2,B) N_2 \over
                      \sum_{N_2=0}^\infty P(R_{\rm exp}|N_2,B)           }\;,
\label{n2av}
\end{equation}
where $P(R_{\rm exp}|N_2,B)$ is given by eq.~(\ref{probr}), and we have
implicitly assumed a uniform prior distribution for $N_2$;
in the present case, eq.~(\ref{n2av}) yields $\langle N_2\rangle = 12$.

Turning to future experiments, we first recall that
BATSE happened to produce a low value of $R_{\exp}$.
A larger catalog for BATSE  which instead yielded the most likely value 
when $f=0$, $R_{\exp}=\langle R\rangle_{f=0}=\frac14 B\sigma^2$, 
would result in
$P(R_{\rm exp}|f)\propto\exp[-(0.15\,f)^2/2(\Delta R)^2]$ for $f>0$.
Assuming (for example) a catalog with $B=3000$ bursts,
we find that $H>3.7\times 10^5$ at the 90\% CL.
More significant improvements in the lower bound on $H$
would require an instrument with a better angular resolution.  

We now turn to the more general case of a population of hosts, each with an 
individual $A_i$ (the burst production rate times observation time for that 
host).   We will make the mild assumption that the distribution of $A_i$'s
has moments
\begin{equation}
{1\over H} \sum_{i=1}^H A_i^n \equiv c_n A^n
\label{mom}
\end{equation}
which remain finite in the limit of large $H$.  
Note that $c_1=1$ and $c_2\ge 1$.  From here on we will concentrate 
on the simplest case of no repetitions.

The probability that we see $B$ bursts with no repetitions is given by
eq.~(\ref{probno}).  We can write the Kronecker symbol as 
\begin{equation}
\delta_{B_1+\ldots+B_H,B} = \oint {dz\over 2\pi i z} \, 
    z^{-B}z^{B_1}\ldots z^{B_H} \, ,
\label{kron}
\end{equation}
where the contour encloses the origin.  Each sum in eq.~(\ref{probno})
now yields a factor of $\exp(-A_i)(1+zA_i)$, and so we find
\begin{eqnarray}
P(B,\hbox{no rep}|\{A_i\},H) 
&=& \exp(-AH) \oint {dz\over 2\pi i z} \, 
    z^{-B} \prod_{i=1}^H(1+z A_i) \nonumber \\
&=& \oint {dz\over 2\pi i z} \, 
    \exp\biggl[\sum_{i=1}^H\log(1+z A_i)-B\log z - AH \biggr] \, . 
\label{probno3}
\end{eqnarray}
We now treat both $H$ and $B$ as large, and evaluate the integral by steepest 
descent.  Let $F(z)$ be the argument of the exponential in the second
line of eq.~(\ref{probno3}); the point $z_0$ of steepest descent 
is then given by $F'(z_0)=0$, or
\begin{eqnarray}
0 &=& \sum_{i=1}^H{A_i\over 1+z_0 A_i}-{B\over z_0} \nonumber \\
  &=& H\bigl[A - z_0(c_2 A^2) + z_0^2(c_3 A^3) + \ldots\bigr] - B z_0^{-1} \, .
\label{steep}
\end{eqnarray}
Taking $B \ll H$, this can be solved by power series to yield
\begin{equation}
z_0 A = (B/H) + c_2(B/H)^2 + (2c_2^2-c_3)(B/H)^3 + \ldots \, .
\label{z0}
\end{equation}
The probability of getting $B$ bursts with no repetitions is now given by
\begin{equation}
P(B,\hbox{no rep}|\{A_i\},H) = [2\pi F''(z_0)]^{-1/2}z_0^{-1}\exp F(z_0) \, .
\label{steep2}
\end{equation}
After dividing by $P(B|A,H)$ we find that the probability that 
no repetitions are seen, given that $B$ bursts are observed, is
\begin{equation}
P(\hbox{no rep}|B,\{A_i\},H) = \exp(-c_2 B^2/2H)
\label{steep3}
\end{equation}
in the limit $H\gg B\gg 1$.  This is qualitatively the same as our previous 
result, eq.~(\ref{ratio2}), except that now the relative value of the second 
moment of the distribution enters as well.  The coefficient $c_2$ must be 
greater than one, so this correction can only increase the lower bound on $H$.

\section{Implications for the hosts of GRBs and Discussion}
 
We now turn to the implications of the preceding analysis to the
specific case of GRBs and the host population in which the GRBs
may be embedded.  We have seen that if we observe
$B$ GRBs, which originated from a host population of $H$ hosts,
and further if assume that no host generated more than
one burst, then we can conclude that $H$ must be bounded below by $B^2$,
at a confidence level of roughly 90\%.
If a probability-weighted average over the number of repeaters
is taken instead, then the lower bound on $H$ drops to a number
which is of order $B^2/\langle N_2\rangle$, where $\langle N_2\rangle$
is the expected number of repeaters.  This is given by eq.~(\ref{n2av}),
which would typically yield $\langle N_2\rangle \sim B\sigma$
(and hence $H_{\rm min} \sim B/\sigma$),
where $\sigma$ is the angular resolution of the detector.
Our constraint on $H$ can be significant for models in which the
host population for GRBs consists of exotic objects of any kind,
such as galaxies with some particular morphological feature or other 
specific signature such as a high star formation rate.
Another possibility is to consider clusters of galaxies
as hosts (rather than individual galaxies), provided that
the angular size of the clusters considered is much less than $\sigma$.

If repeaters are ever detected statistically through a value
of $R_{\rm exp}$ which is larger than $\langle R\rangle$
by several $\Delta R$, this analysis can be readily adapted 
to find upper and lower limits on the number of hosts $H$.  
The most likely value will be $H=B^2/\langle N_2\rangle$, 
where $\langle N_2\rangle$
is the most likely value of the number of repeaters.

Finally, we restate our main point: simply by knowing that there
are $\sim 10^3$ bursts with no sign of repeaters, we can immediately
infer a model-independent lower bound on the number of hosts 
in which GRBs may be embedded which is larger by more than two orders
of magnitude.

\acknowledgments

We thank Omer Blaes for many helpful discussions, and the referee
for several important comments which improved the paper.
This work was supported in part by NSF Grant PHY--91--16964.


\begin{references}

\reference{Omer94} Blaes, O. 1994, \apjs, 92, 643

\reference{Omeretal97} Blaes, O., Hurt, T., Antonucci, R., Hurley, K. \&
Smette, A. 1997, \apj, 479, 868

\reference{Fenimoretal} Fenimore, E.E. et al. 1993, \nat, 366, 40

\reference{Hurleyetal} Hurley, K., Hartmann, D., Kouveliotou, C., Fishman, G.,
Laros, J., Cline, T., and Boer, M. 1997, \apj, in press.

\reference{KP96} Kollat, T. \& Piran T. 1996, \apjl, 467, L41

\reference{Larson} Larson, S., Mc Lean, I. \& Becklin, E. 1996,
\apjl, 460, L95

\reference{Loredo} Loredo, T.J., and Wasserman, I. 1995,
\apjs, 96, 261

\reference{Metzgeretal} Metzger, M., Djorgovski, S., Steidel, C., Kulkarni, S.,
Adelberger, K. \& Frail, D. 1997, IAUC 6655.

\reference{NarayanPiran} Narayan, R. \& Piran, T. 1993, \mnras, 265, L65

\reference{Pet93} Petrosian V. 1993, \apjl, 402, L33

\reference{Pi95} Piran, T. 1995, in 
Some unsolved problems in astrophysics, 
eds. Bahcall, J. \& Ostriker, J. (Princeton: Princeton Univ. Press).

\reference{PS96} Piran, T. \& Singh, A. 1997, preprint astro-ph/9607072,
to appear in \apj, July 1997

\reference{LQ} Quashnock, J. \&  Lamb, D. 1993, \mnras, 265, L59

\reference{Scaeferetal} Schaefer, B., Cline, T., Hurley, K. \& Laros, G.
1997, preprint astro-ph/9704278.

\reference{Tegmarketal} Tegmark, M., Hartmann, D., Briggs, M., Hakkila, J.
\& Meegan, C. 1996, \apj, 466, 757

\end{references}
\end{document}